\documentclass[sigconf, natbib=false]{acmart}

\AtBeginDocument{%
  }

\setcopyright{acmcopyright}
\copyrightyear{2024}
\acmYear{2024}
\acmDOI{XXXXXXX.XXXXXXX}

\usepackage{enumitem}
\setlist[itemize]{noitemsep, topsep=0pt}

\newcommand{\cs}[1]{\texttt{#1}}

\newcommand{\floor}[1]{\left\lfloor #1 \right\rfloor}

\usepackage{listings}
\definecolor{codegreen}{rgb}{0,0.6,0}
\lstdefinestyle{customcpp}{
    breakatwhitespace=true,
    breaklines=true,
    frame=single,
    xleftmargin=\parindent,
    language=C++,
    showstringspaces=false,
    basicstyle=\footnotesize\ttfamily,
    keywordstyle=\bfseries\color{blue!80},
    commentstyle=\itshape\color{white!40!black},
    identifierstyle=\color{black},
    stringstyle=\color{green!60!black},
    backgroundcolor=\color{white},
    tabsize=2,
    showspaces=false,
    firstnumber=1,
}

\RequirePackage[
  datamodel=acmdatamodel,
  style=acmnumeric,
  ]{biblatex}

\begin{document}

%%
%% The "title" command has an optional parameter,
%% allowing the author to define a "short title" to be used in page headers.
\title[Lessons Learned Migrating CUDA to SYCL]{Lessons Learned Migrating CUDA to SYCL:\\ A HEP Case Study with ROOT RDataFrame}

%%
%% The "author" command and its associated commands are used to define
%% the authors and their affiliations.
%% Of note is the shared affiliation of the first two authors, and the
%% "authornote" and "authornotemark" commands
%% used to denote shared contribution to the research.
\author{Jolly Chen}
\email{jolly.chen@cern.ch}
\orcid{0000-0003-1110-1256}
% \authornotemark[1]
\affiliation{%
\institution{University of Amsterdam}
  \institution{CERN}
  \city{Geneva}
  \country{Switzerland}
}

\author{Monica Dessole}
\affiliation{%
  \institution{CERN}
  \streetaddress{}
  \city{Geneva}
  \country{Switzerland}}
\email{monica.dessole@cern.ch}

\author{Ana Lucia Varbanescu}
\affiliation{%
  \institution{University of Twente}
  \city{Enschede}
  \country{The Netherlands}}
\email{a.l.varbanescu@utwente.nl}

%%
%% By default, the full list of authors will be used in the page
%% headers. Often, this list is too long, and will overlap
%% other information printed in the page headers. This command allows
%% the author to define a more concise list
%% of authors' names for this purpose.
\renewcommand{\shortauthors}{Chen et al.}

%%
%% The abstract is a short summary of the work to be presented in the
%% article.
\begin{abstract}
The world's largest particle accelerator, located at CERN, produces petabytes of data that need to be analysed efficiently, to study the fundamental structures of our universe. ROOT is an open-source C++ data analysis framework, developed for this purpose. Its high-level data analysis interface, RDataFrame, currently only supports CPU parallelism. Given the increasing heterogeneity in computing facilities, it becomes crucial to efficiently support GPGPUs to take advantage of the available resources. SYCL allows for a single-source implementation, which enables support for different architectures.
In this paper, we describe a CUDA implementation and the migration process to SYCL, focusing on a core high energy physics operation in RDataFrame -- histogramming. We detail the challenges that we faced when integrating SYCL into a large and complex code base. Furthermore, we perform an extensive comparative performance analysis of two SYCL compilers, AdaptiveCpp and DPC++, and the reference CUDA implementation. 
We highlight the performance bottlenecks that we encountered, and the methodology used to detect these. Based on our findings, we provide actionable insights for developers of SYCL applications.

\end{abstract}

%%
%% The code below is generated by the tool at http://dl.acm.org/ccs.cfm.
%% Please copy and paste the code instead of the example below.
%%


%%
%% Keywords. The author(s) should pick words that accurately describe
%% the work being presented. Separate the keywords with commas.

% \received{19 January 2023}
% \received[revised]{12 March 2009}
% \received[accepted]{5 June 2009}

%%
%% This command processes the author and affiliation and title
%% information and builds the first part of the formatted document.
\maketitle

\section{Introduction}
CERN hosts the Large Hadron Collider (LHC), the world's largest particle accelerator, used by physicists all over the world to empirically study the fundamental structures of our universe. For all LHC experiments, millions of particle collisions happen inside the accelerator every second, resulting in petabytes of data that need to be analysed \cite{cern_computing_storage}. These collisions are recorded in a digitised format we refer to as \textit{events}. With the upcoming upgrade to the High Luminosity LHC, the intensity is expected to increase by a factor of 30 \cite{hep2019roadmap}, leading to a similar increase in data to be analysed.

Analysing these massive amounts of data naturally requires highly efficient and complex software solutions. ROOT \cite{root} is an open-source platform-independent C++ data analysis framework designed and developed for this purpose. The ROOT project provides the core ingredients for high energy physics (HEP) analysis tasks, including optimised data storage, a user-friendly and interactive interface, math libraries, and visualisation capabilities. 

One of the main components of ROOT is \textit{RDataFrame} \cite{rdf_paper}, the high-level data analysis interface. 
\texttt{RDataFrame} represents physics data in a columnar format, where a row defines a collision event and the columns describe various characteristics for each event. A HEP analysis generally consists of iterating over the data from different events to compute distributions (histogramming), apply filters, and derive new computed columns. Within ROOT, these operations must be performed efficiently, thus making use of modern hardware such as parallel CPUs and accelerators is an implicit requirement.
To this end, one of the main design goals of the RDataFrame interface is to have easy-to-enable parallelism and portability, without  requiring extensive knowledge of parallel computing and/or programming from its users. For example, enabling parallel execution can be as simple as adding the line \texttt{ROOT::EnableImplicitMT()}~\footnote{This call enables multi-threading (MT) for multi-core CPUs.} and requires no further modifications to the code. Currently, RDataFrame contains support for implicit parallelism in multi-threaded \cite{rdf_paper} and multi-node distributed environments \cite{rdf_distributed}, but GPU parallelism has not been attempted before this work. 

The upcoming demand for more computing power, paired with the trend of increased heterogeneity in computing facilities, makes it imperative that efforts are put into developing software capable of targeting the full system. Our initial approach, using CUDA, demonstrates relevant performance gains from using accelerators, but adds too much complexity to the software by requiring maintenance of two different code bases. Instead, in this work, we investigate the use of SYCL for enabling heterogeneous computing within RDataFrame. We used the two most mature SYCL implementations currently available, Intel's DPC++ \cite{dpcpp_github} and AdaptiveCpp (formerly hipSYCL/OpenSYCL) \cite{acpp}. In our CUDA developments, we started with offloading the histogram action, one of the core HEP operations, and the most suitable for acceleration. By re-implementing this core operation with SYCL, we can enable support for architectures beyond NVIDIA GPUs with a single-source implementation. This paper describes the lessons learned in this process, for the specifics of histogramming with \texttt{RDataFrame}. 

Specifically, we make the following contributions:
\begin{itemize}[leftmargin=*]
    \item We present our process for migrating existing CUDA code to SYCL, within a large and complex C++ code base. We focus on the histogram action, one of the most commonly used actions in RDataFrame. 
    \item We detail the various performance and compatibility challenges we have encountered, a methodology for detecting them, and several solutions to tackle them. This includes a comparison between SYCL buffers and device pointers, work-item organisation for reductions, and just-in-time compilation caching.
    \item We provide an extensive comparative performance analysis of GPU-based histograms using two SYCL implementations - AdaptiveCpp and Intel's DPC++ - and a native CUDA implementation. 
\end{itemize}
\vspace{3pt}

The remainder of this paper is organised as follows: in \autoref{sec:background} we briefly describe the usage and inner workings of RDataFrame. In \autoref{sec:implementation}, we describe the implementation of the ROOT components that have been ported to the GPU. In \autoref{sec:evaluation} we evaluate the performance of our implementations. We present our method, experimental setup, a detailed comparative performance analysis, and performance debugging suggestions. In \autoref{sec:related}, we discuss and compare our work in the context of alternative SYCL studies. Finally, \autoref{sec:conclusion} summarises our findings and sketches future work directions. 

\section{Background}\label{sec:background}
\begin{figure}
    \centering
    \includegraphics[width=0.7\columnwidth]{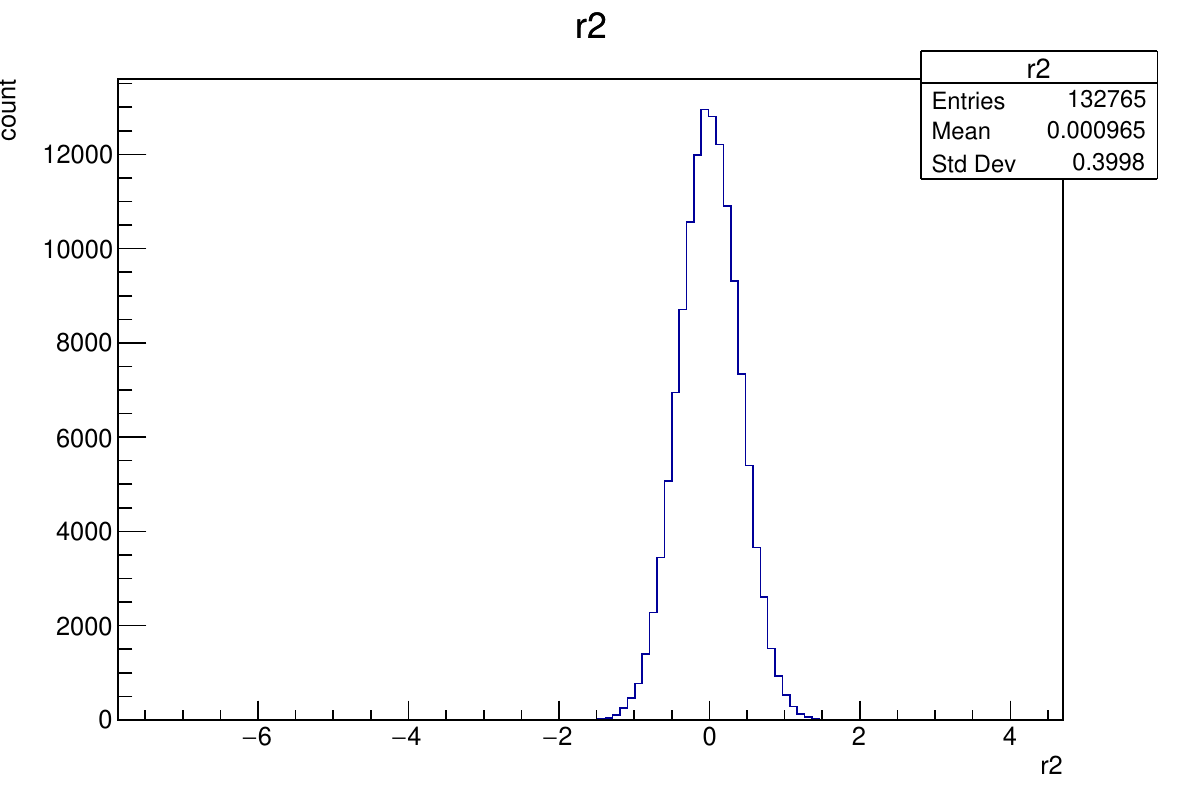}
    \caption{Example of a ROOT 1D histogram. }
    \label{fig:example_histogram}
\end{figure}

A simple example of the RDataFrame interface is shown in \autoref{lst:df000_simple}. To analyse the data, the user can book various actions that are applied to each event. In this example, the user fills a one-dimensional histogram. On the first line of the code example, a data frame is created from a data source, which is commonly a compressed data format such as a ROOT TTree file \cite{ttree} or an RNTuple file (the next-generation TTree) \cite{rntuple}. After the creation of the data frame, two actions are booked: filtering rows where the value in column \texttt{x} is non-zero and defining a new column \texttt{r2} based on the values in column \texttt{x} and \texttt{y}. After that, a histogram action is booked, where a 1D histogram is filled with the values in column \texttt{r2}. Lastly, the result of the previous actions is visualized by drawing the histogram (e.g., \autoref{fig:example_histogram}), which triggers the computational loop over the rows in the data frame. 
\begin{lstlisting}[language=c++, style=customcpp, label=lst:df000_simple, caption=Simple RDataFrame example in C++][H]
ROOT::RDataFrame df(data_source, bulksize)
auto df = df.Filter("x != 0")
            .Define("r2", "x*x + y*y")
auto h = df.Histo1D<double>("r2")
h->Draw();
\end{lstlisting}

One of the main features of RDataFrame is the lazy execution of the actions. In the previous example, the filter, define, and histogram actions are processed in sequence per event in a single loop over all the events in the dataframe. Each event is in principle independent of other events, which allows for parallelisation over the events. By defining an internal computational graph, the framework can execute several paths and combinations of actions simultaneously. 

In the current ROOT release (6.31), the actions are processed event-by-event, but recent developments include bulk-by-bulk processing of events \cite{rdf_bulk_branch}. This provides a natural unit of data to offload to accelerators. The implicit parallelism mentioned previously is across bulks, but parallelism within bulks has not been implemented yet. In this paper, we based our work on RDataFrame with the bulk API. 

\section{Implementation}\label{sec:implementation}
In this section, we describe the implementation of the histogram action within ROOT in more detail. In \autoref{sec:cpu-implementation}, we first describe which computations are needed for the histogramming action. In \autoref{sec:cuda-implementation}, we go into our native CUDA implementation and in \autoref{sec:sycl-implementation} we describe our process of porting the CUDA code to SYCL.

\begin{figure}
    \centering
    \includegraphics[width=\columnwidth]{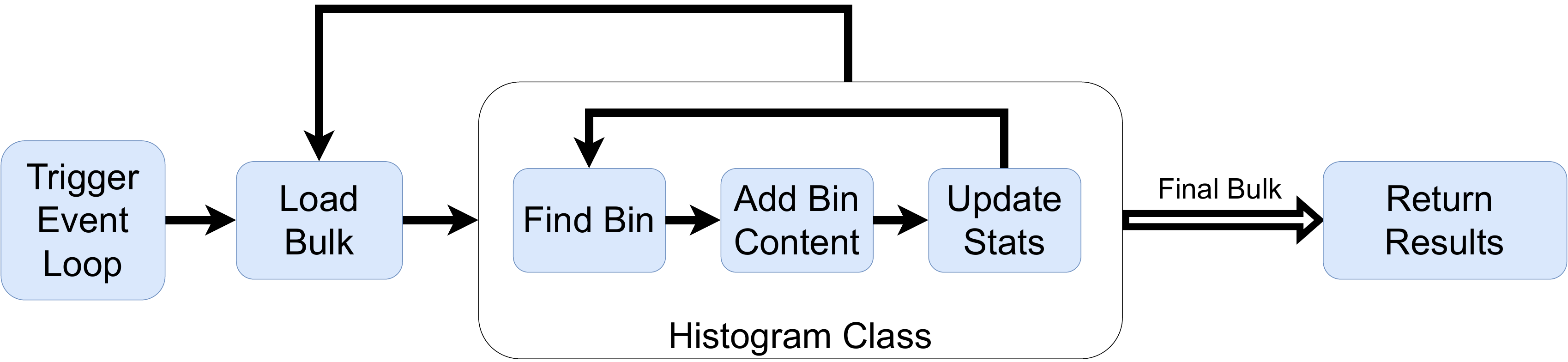}
    \caption{Processing of a histogram action in RDataFrame.}
    \label{fig:histond-action}
\end{figure}

\subsection{Histogramming Action}\label{sec:cpu-implementation}
When a computational loop in RDataFrame is triggered, we iterate over the events in the dataframe and determine which actions need to be processed for that event. In the case of the histogram action, the class \texttt{TH1} represents a 1D histogram and implements the \texttt{Fill} method. The \texttt{FillHelper} class, which is initialised at the beginning of the loop, keeps a reference to the \texttt{TH1} class and passes bulks of events to the \texttt{Fill} method. 

As illustrated in \autoref{fig:histond-action}, the \texttt{Fill} method for a bulk of events involves three main steps: (1) determining which bin needs to be filled based on an input coordinate, (2) incrementing said bin with a given weight, and (3) updating sum variables for histogram statistics (e.g., the mean, standard deviation, and higher momenta). The steps required to find the bin depend on the histogram's edges, which can be fixed-sized or variable-sized. In the first case, the appropriate bin can be computed via a simple formula with constant time (i.e., $1+\floor{nbins*\frac{coord-minEdge}{maxEdge-minEdge}} $). For variable-sized bins, however, a search algorithm needs to be used. In ROOT, we use a binary search, implemented using \texttt{std::lower\_bound} on the CPU. Additionally, in the \texttt{FindBin} method, we check whether the coordinate falls outside of the histogram's range, in which case a dedicated underflow or overflow bin is filled instead. Once the bin is determined, the bin content is incremented by one or a given weight. For the reduction, the number of statistics depends on the dimension of the histogram. Higher dimension histograms also require repeating the ``find bin" step per axis. 

\subsection{Native CUDA}\label{sec:cuda-implementation}
\begin{figure}[t]
    \centering
    \includegraphics[width=\columnwidth]{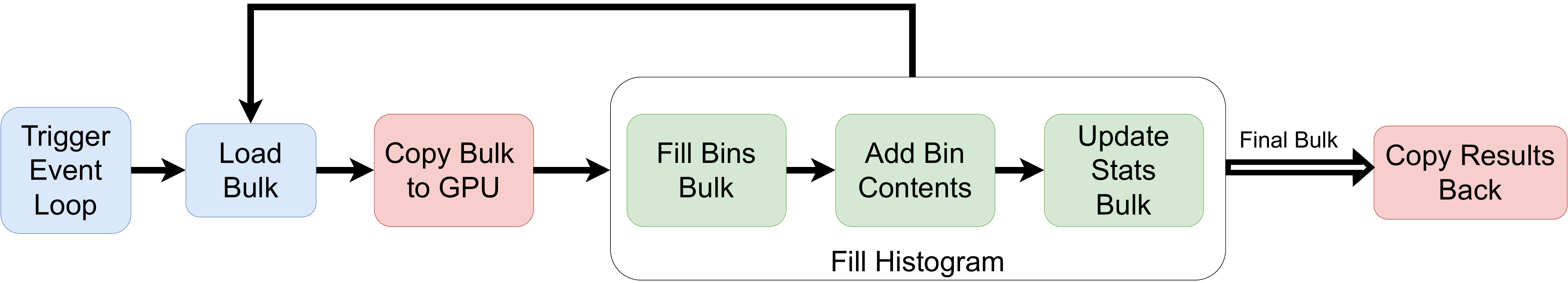}
    \caption{Processing of a histogram action in RDataFrame with a GPU. Red = memory transfers, green = GPU execution, and blue = CPU execution.}
    \label{fig:histond-action-gpu}
\end{figure}

Enabling heterogeneous computing in RDataFrame involves two parts: (i) create/modify action processing methods to execute on a GPU and (ii) implement a helper class that passes a bulk of input events to the responsible action methods while keeping track of the results. For the native CUDA version, we implemented a \texttt{RHnCUDA} histogram class that replaces the \texttt{TH1} class, and a matching \texttt{CUDAFillHelper} that replaces the CPU \texttt{FillHelper}. The \texttt{RHnCUDA} class keeps track of device allocations and contains a \texttt{Fill} method. This transfers bulk of events to the GPU and launches kernels to perform histogram computations per bulk. To trigger the CUDA code path instead of the CPU path, the user runs the same RDataFrame code but with the \texttt{CUDA\_HIST} environment variable set, so that the \texttt{CUDAFillHelper} is selected instead. The modified flow is illustrated in \autoref{fig:histond-action-gpu}. While iterating over the events in our dataframe, we transfer bulks of events to process them on the GPU. We only copy back the resulting histogram once all bulks have been processed.  

We implemented two base kernels that compute the three histogram steps: a histogram kernel and a reduction kernel. The histogram kernel is displayed in \autoref{lst:cuda-histogram}, and shows how we perform the filling in two steps: (1) each block fills a local copy of the histogram stored in shared memory using a block-wide atomic operation, and (2) once the local histograms are filled, the results are combined into the global histogram stored in global memory using a grid-wide atomic. By filling in two stages, we reduce the potential atomic contention. Since C++ standard library functions are not available with CUDA, the \texttt{std::lower\_bound} call in the binary search for the bin retrieval is replaced with \texttt{thrust::lower\_bound}. 

\begin{lstlisting}[float=htpb,language=c++, style=customcpp, label=lst:cuda-histogram, caption=CUDA Histogram kernel][h]
template <typename T, unsigned int Dim>
__global__ void HistogramLocal(...) {
  auto sMem = init_smem<T>();
  auto tid = threadIdx.x + blockDim.x * blockIdx.x;
  auto localTid = threadIdx.x;
  auto stride = blockDim.x * gridDim.x; 

  // Initialize a local per-block histogram
  for (auto i=localTid; i<nBins; i+=blockDim.x) {
    sMem[i] = 0;
  }
  __syncthreads();

  // Fill local histogram
  for (auto i=tid; i<bulkSize; i+=stride) {
    auto bin = GetBin<Dim>(...);
    if (bin >= 0) 
      atomicAdd(&sMem[bin], (T)weight);
  }
  __syncthreads();

  // Merge results in global histogram
  for (auto i=localTid; i<nBins; i+=blockDim.x) {
    atomicAdd(&histogram[bin], (T)weight);
}}
\end{lstlisting}

The computations required for the histogram statistics are essentially \textit{transform-reduce} operations, i.e.,  operations that transform each value in a bulk with some operator, and then produce a single value by combining all values in a bulk with a binary operator. In the one-dimensional histogram case, we have four reductions: the sum of (squared) weights and the sum of (squared) weights multiplied by the corresponding coordinate. The implementation of the reduction is based on the sample kernel provided by CUDA \cite{reduction_sample}, which reduces multiple elements per thread sequentially, modified to accept a custom initialization operator and combinator operation as template arguments. 

% Strategy porting to sycl
\subsection{Porting to SYCL}\label{sec:sycl-implementation}
We ported each CUDA kernel to named SYCL function objects with identical behaviour. Starting with the histogram kernel, shown in \autoref{lst:sycl-histogram}, the original CUDA code could be translated almost one-to-one. While the SYCL kernel encompasses more lines of code than the CUDA version at first sight, this is mostly due to long namespaces. For the statistics, we used the reduction interface defined in the SYCL2020 standard \cite{sycl2020}. 

The SYCL code remains unchanged when switching between AdaptiveCpp and DPC++, except for the reduction part. For DPC++, the built-in reduction kernel keeps the original value of the reduction variable and combines it with the reduced result. With AdaptiveCpp, however, the reduction variable is overwritten. We noticed this when we switched from DPC++ to AdaptiveCpp and got incorrect results. The current behaviour of AdaptiveCpp contradicts the SYCL2020 specification, which states: `\textit{The initial value of the reduction variable is included in the reduction operation unless the \texttt{property::reduction::initialize\_to\_identity} property was specified}' \cite{sycl2020}. Since the SYCL compilers are still under active development, missing or incomplete features are to be expected. 
In our case, we desire the specified behaviour, as the sum of the previous bulk needs to be combined with the result of the current bulk. Therefore, in the AdaptiveCpp version, we launch an extra kernel that iterates over the statistics array to combine the results. 

\begin{lstlisting}[float=htpb, language=c++, style=customcpp, label=lst:sycl-histogram, caption=SYCL Histogram kernel][t]
template <typename T, unsigned int Dim>
class Histogram {
  void operator()(sycl::nd_item<1> item) const {
    auto globalId = item.get_global_id(0);
    auto localId = item.get_local_id(0);
    auto group = item.get_group();
    auto groupSize = item.get_local_range(0);
    auto stride = groupSize * item.get_group_range(0);

    // Initialize a local per-work-group histogram
    for (auto i=localId; i<nBins; i+=groupSize) {
      localMem[i] = 0;
    }
    sycl::group_barrier(group);

    // Fill local histogram
    for (auto i=globalId; i<bulkSize; i+=stride) {
      auto bin = GetBin<Dim>(...);

      if (bin >= 0) {
        auto atomic =
          sycl::atomic_ref<T,
            sycl::memory_order::relaxed,           
            sycl::memory_scope::device,           
            sycl::access::address_space::local_space>(
              localMem[bin]);
        atomic.fetch_add(weight);
      }
    }
    sycl::group_barrier(group);

    // Merge results in global histogram
    for (auto i=localId; i<nBins; i+=groupSize) {
      auto atomic =
          sycl::atomic_ref<T, 
            sycl::memory_order::relaxed, 
            sycl::memory_scope::device, 
            sycl::access::address_space::global_space>(
              histogram[bin]);
      atomic.fetch_add(weight);
}}};
\end{lstlisting}

% cmake
One of the main difficulties of adding SYCL within our code base was the integration of a SYCL implementation's compilation flow into ROOT's CMake build system. AdaptiveCpp provides a CMake function \texttt{add\_sycl\_to\_target } that allows for compiling only the SYCL files with their \texttt{syclcc} compiler. On the other hand, the \texttt{IntelDPCPPConfig.cmake} configuration package provided with DPC++ requires setting the definition of \texttt{CMAKE\_CXX\_COMPILER}, so the entire project is built with their SYCL compiler. However, ROOT's built-in interactive C++ interpreter, Cling \cite{cling}, relies on some custom patches to LLVM, so this option easily leads to incompatibilities. Moreover, modifying the \texttt{CMAKE\_CXX\_COMPILER} string for only specific targets is not allowed by CMake. To ensure that only our SYCL code is compiled with the DPC++ compiler, we used CMake's \texttt{add\_custom\_command} to override the original build and linking rule. In more detail, we implemented a CMake function \texttt{add\_sycl\_to\_root\_target} that takes a CMake target and creates a custom command for compiling each source file and linking the target with the necessary flags, based on the target's properties. With this method, CMake targets with SYCL files can be created and modified with the usual CMake commands (e.g., \texttt{target\_include\_directories} and \texttt{target\_link\_libraries}) and adding the SYCL compilation flow only requires calling the function \texttt{add\_sycl\_to\_root\_target} for the target. Lastly, we also enable toggling between two different SYCL implementations defining this function to either call on AdaptiveCpp's \texttt{add\_sycl\_to\_target} or to apply the custom commands for DPC++, based on a CMake configuration flag. 

An added benefit of compiling our code with two different SYCL compilers was that it helped us discover bugs based on undefined behaviour in our code. For example, the SYCL specification does not specify when enqueued work is executed, so an implementation could defer execution until a batch of operations can be processed. As a result, some variables might not be in scope anymore which could lead to incorrect results without prior synchronization. In our case, we have bulks of event data that need to be copied one after another to the GPU. However, the host container was occasionally overwritten with data from the next bulk, before the asynchronous copy of the current bulk had completed. To fix this, we had to add a synchronization point after the bulk computations and before the transfer of a new bulk. Note that this only occurred in AdaptiveCpp when we used SYCL's USM device pointers, which we discuss more in \autoref{sec:buf_vs_ptr}.

\section{Evaluation} \label{sec:evaluation}
In this section, we present a detailed comparative analysis of three accelerated 1D histograms within RDataFrame: the CUDA version, used as reference, and the two SYCL versions - AdaptiveCpp and DPC++. We focus on the three biggest performance bottlenecks we discovered during the migration of our code, and highlight their impact on overall SYCL code porting/development. Specifically, (1) we compare the performance of different SYCL2020 reduction implementations, and (2) we perform a buffers vs. device pointers analysis. In addition, (3) we emphasise the importance of target specification and just-in-time compilation caching. The first two topics influence the implementation, where the first one requires more effort than the second, while the third one concerns the build configuration of the application. 

The remainder of this section is structured as follows: in \autoref{sec:analysis-method}, we describe the methods we used to measure the performance of our application and discover performance bottlenecks. In \autoref{sec:experimental-setup}, we outline our test environment and the benchmark that we used. In sections \ref{sec:reduction}, \ref{sec:buf_vs_ptr}, and \ref{sec:aot_target} we discuss the aforementioned performance bottlenecks.

\subsection{Evaluation method}\label{sec:analysis-method}
We used the following methods to analyse the performance of different versions of our program.

\textbf{Method 1. Measure total runtime using a wall-clock time}. The most straightforward way to acquire runtime information, with minimal overhead, is to measure the time it takes to execute the full program or the relevant section from start to end in real-time. The most suitable clock for measuring intervals, according to the C++ reference, is \texttt{std::chrono::steady\_clock}. It is a monotonic clock, which guarantees that the clock time does not decrease when the physical time increases and the time between clock ticks remains constant. 

\textbf{Method 2. Run the program with a profiler to determine the amount of time spent on specific GPU activities}. While method 1 gives us an insight into the total performance, we require information at a finer granularity to identify the source of GPU performance changes. In this work, we use NSight Systems version 2023.4.1.97-234133557503v0 which can measure time spent in \textit{CUDA API calls}, \textit{kernels}, and \textit{memory operations}. We show that performance data from these three classes of operations provide insight into the internal behaviour of the SYCL implementations. The values we inspect are total time, number of calls/instances, and kernel launch parameters. 

\subsection{Experimental Setup}\label{sec:experimental-setup}
We run our benchmarks on a single GPU node on the DAS-6 cluster \cite{das6}. The node is equipped with an AMD EPYC 7402P 24-core Processor, 128 GB of main memory, and an NVIDIA RTX A4000 GPU with driver version 545.23.06 and compute capability 8.6 (Ampere). The host code is compiled using version 12.2.1 of the GCC compiler, while the files containing device code are compiled using \cs{nvcc} from the CUDA Toolkit version 12.3, the \cs{syclcc} compiler of AdaptiveCpp, and \cs{clang++} of Intel OneAPI's DPC++. Since both SYCL compilers are still being actively developed, we build both AdaptiveCpp \cite{acpp_github} (commit 67cb7a) and DPC++ \cite{dpcpp_github} (commit dbee22) from source to get access to the latest developments. AdaptiveCpp was built with LLVM 16.0.6 and DPC++ was built with OpenCL 2023.16.6.0.28. The ROOT code is compiled using \texttt{-O3}, and we used the multi-pass compilation flow for AdaptiveCpp \cite{acpp_compilation}. Unlike DPC++, AdaptiveCpp also provides a single-pass compiler that parses both host and device code at once, which promises a higher degree of portability, low compilation times, flexibility and extensibility. However, SYCL2020 reductions are not yet supported in this compilation mode. 

\begin{lstlisting}[language=c++, style=customcpp, label=lst:histond_benchmark, caption=Benchmarking application pseudocode][h]  
RDataFrame df(data_source, /* bulksize */ 32768);
auto edges = generate_edges() // nbins in range [0,1]
auto mdl = TH1DModel("h1", "h1", nbins, edges));

auto start = Clock::now();
auto h = df.Histo1D<double>(mdl, "columnName");
auto &result = h.GetValue();
auto end = Clock::now();
\end{lstlisting}

For benchmarking, we measure the execution time of \texttt{Histo1D} (pseudocode shown in \autoref{lst:histond_benchmark}), since our current implementations only execute the RDataFrame histogramming action on the GPU. As a representative histogram, we choose to use a 1D histogram with 1000 bins in the range [0.0, 1.0]. Given that only the number of bins has an impact on the bin search time, we selected arbitrary values for the range extremes. Furthermore, although real-world RNTuple datasets commonly contain a multitude of fields/columns with different data types, we focus in our analysis on synthetic datasets with a single data type: double-precision floating point ("double") - the commonly used data type for histograms. Therefore, the input consists of four RNTuples, each containing a single column per event, with 50M, 100M, 500M, or 1B doubles in total\footnote{M = million, B = billion}; this setup is generic and large enough, but also limits the total runtime to tens of seconds. The values are uniformly distributed in the range [0,1], which exercises filling the full histogram. While other distributions may influence the bin search and filling time, we leave this analysis (and potential performance impacts/improvements for such distributions) outside this work, and focus on our main objective: a detailed analysis of CUDA vs. SYCL.

%%%%%%%%%%%%%%%%%%%%%%%%%%%%%%%%%%%%%%%%%%%%%%%%%%%%%%%%%%%%%%%%%%%%%%%%%%%%%%%%%%%%%%%%%%%%%%%%%%%%%%%%%%%%%%%%

\subsection{SYCL2020 Reduction}\label{sec:reduction}
\noindent\fcolorbox{black}{ACMBlue!30}{
    \parbox{0.95\columnwidth}{%
\textbf{Insights gained:}
\begin{itemize}[leftmargin=*]
    \item Increasing the workload per work-item in the SYCL2020 reduction kernel greatly improves the performance of AdaptiveCpp's reduction.
    \item Combining multiple reduction variables in a reduction kernel significantly improves the reduction performance for both AdaptiveCpp and DPC++.
\end{itemize}
}}
\newline

Since the \textit{transform-reduce} operation is a common programming pattern, SYCL 2020 provides a reduction interface that simplifies expressing reduction semantics using SYCL kernels. A basic example is shown in \autoref{lst:reduction_basic}. First, the user defines a \textit{reduction variable} using the \texttt{sycl::reduction} class. This class accepts a SYCL buffer or an USM pointer as input container and a combinator (\textit{reduce}) operator. The reduction variable is then passed to a \texttt{parallel\_for} construct together with a reduction function. Within this lambda function, the user can define the \textit{transformation} of the input value before the reduction. For example, \texttt{sum += inputValues[idx]} can be replaced with \texttt{sum += inputValues[idx] * inputValues[idx]} to compute the squared sum instead. Note that the \texttt{parallel\_for} is constructed with a single-dimensional range, which leaves the selection of the best work-group/block size to the SYCL implementation. 

\begin{lstlisting}[language=c++, style=customcpp, label=lst:reduction_basic, caption=Basic SYCL2020 Reduction example][t]
myQueue.submit([&](handler& cgh) {
  auto inputValues = 
    valuesBuf.get_access<sycl::access_mode::read>(
        cgh);
  auto GetSum = sycl::reduction(sumBuf, cgh,
    sycl::plus<>());
    
  cgh.parallel_for(range<1> { 1024 }, GetSum,
    [=](id<1> idx, auto& sum) {
        sum += inputValues[idx];
    });
});
\end{lstlisting}

In the reduction kernels of our CUDA version, each thread sequentially sums multiple elements into shared memory before doing a tree-based reduction. This amortises the latency of copying input values into the shared memory. In our SYCL kernels, we can also benefit from increasing the sequential work per work-item. Similarly to our CUDA implementation, we can do this by reducing the number of work-items (the range) and by adding a grid-stride loop as shown in \autoref{lst:reduction_melem_item}.

\begin{lstlisting}[language=c++, style=customcpp, label=lst:reduction_melem_item, caption=SYCL2020 Reduction example with multiple elements per work-item][t]
myQueue.submit([&](handler& cgh) {
  auto inputValues = 
    valuesBuf.get_access<sycl::access_mode::read>(
        cgh);
  auto GetSum = sycl::reduction(sumBuf, cgh,
    sycl::plus<>());
  auto subrange = sycl::range<1>(1024/n)
  cgh.parallel_for(subrange, GetSum,
    [=](sycl::id<1> id, auto &sum) {
      for (auto gid = id; gid < 1024; gid += subrange) {
        sum += inputValues[gid];
      }
    });
});
\end{lstlisting}

An optimisation in our SYCL versions over the CUDA version is that multiple statistics are reduced in each kernel, instead of a single reduction per kernel. Since all our reduction kernels read from the same input arrays (the bin coordinates and weights), combining the reduction operations in one kernel could decrease the overhead of reading/writing values from/into the shared memory. In addition, with fewer kernels, we have a lower kernel launch overhead. With SYCL's API, this was as simple as adding another reduction variable to the command group construction, as shown in \autoref{lst:reduction_mvars_kernel}, where we compute both \texttt{sum} and \texttt{max} in a single \texttt{parallel\_for} construct. Implementing this manually in our native CUDA version would have been very time-consuming and error-prone.  

\begin{lstlisting}[language=c++, style=customcpp, label=lst:reduction_mvars_kernel, caption=SYCL2020 Reduction example with multiple reduction variables per kernel][t]
myQueue.submit([&](handler& cgh) {
  auto inputValues = 
    valuesBuf.get_access<sycl::access_mode::read>(
        cgh);
  auto GetSum = sycl::reduction(sumBuf, cgh,
    sycl::plus<>());
  auto GetSum2 = sycl::reduction(sum2Buf, cgh,
    sycl::plus<>());
  auto subrange = sycl::range<1>(1024/n)
  cgh.parallel_for(subrange, GetSum, GetSum2,
    [=](sycl::id<1> id, auto &sum, auto &max) {
      for (auto gid = id; gid < 1024; gid += subrange) {
        sum += inputValues[gid];
        sum2 += inputValues[gid] * inputValues[gid];
      }
    });
});
\end{lstlisting}

\begin{figure}[t]
    \centering
    \includegraphics[width=\columnwidth]{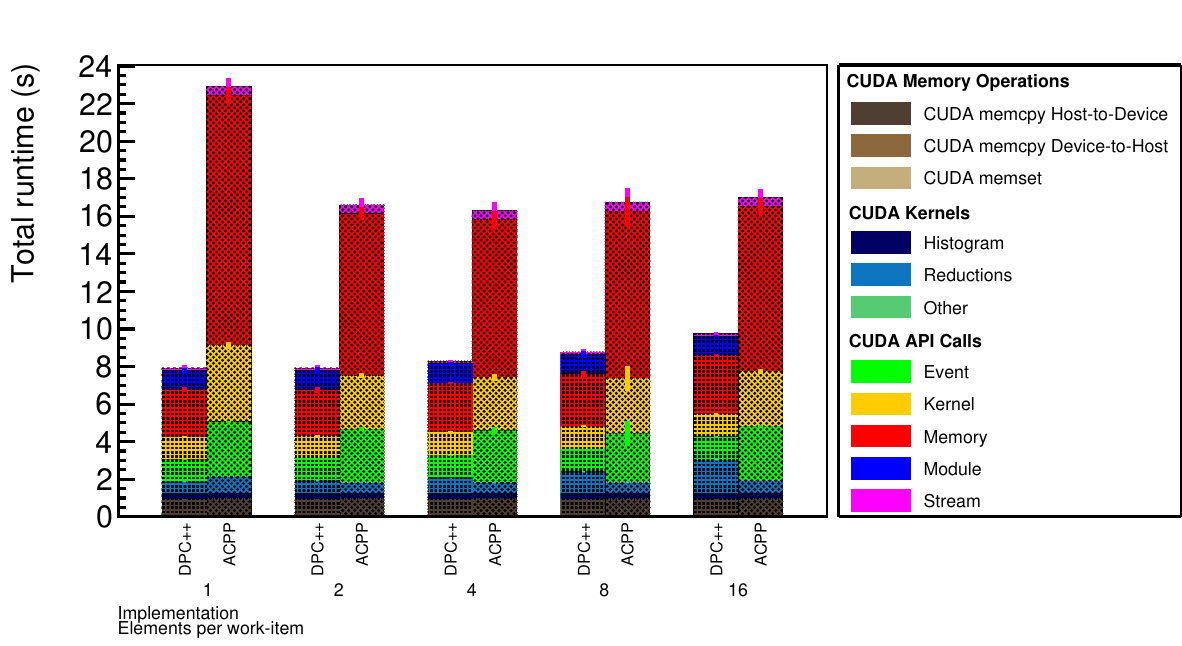}
    \caption{Total time spent on GPU activity in \texttt{Histo1D} with increasing number of elements reduced per work-item using SYCL2020 reductions. Each run processes 1B events in total, with multiple reduction variables in a single SYCL kernel.}
    \label{fig:reduction-range-kernel}
\end{figure}

To show the effect of reducing multiple elements per work-item, we measured the histogram action with one billion events with both setups. \autoref{fig:reduction-range-kernel} shows the profiled GPU activity using Nsight Systems. The graph plots the time spent in kernels, memory operations, and CUDA API calls separately. We group the kernels into \textit{histogram} (steps 1 and 2 from \autoref{sec:cpu-implementation}), \textit{reduction} (step 3), and \textit{other}. The “other'' category describes, for instance, the statistics-combining kernel for the AdaptiveCpp version. Since the CUDA API calls for the two SYCL implementations are named differently, we also combined these into 5 categories for clarity: event, kernel, memory, module, and stream operations. The CUDA memory operations refer to memory transfer time, while the CUDA memory API calls refer to the time spent on setting up the memory operations. The CUDA memory API calls include (de)allocations and copying. Note that the CUDA API calls are executed by the CPU, while the kernels are executed on the GPU, so there is some overlap in runtime that is not illustrated. 

With AdaptiveCpp, we notice a significant difference in execution time; there is an improvement of about 8 seconds (speedup of $\sim$1.4x) with 4 elements per work-item compared to having a single element per work-item. With more elements per work-item, the performance worsens again, because the loss in parallelism outweighs the shared-memory latency-hiding benefit. On the other hand, the difference in runtime for different work sizes is marginal for DPC++. When inspecting the CUDA GPU Kernel/Grid/Block Summary in NSight Systems for the case with two elements per work-item, we notice that AdaptiveCpp used 128 blocks with 128 threads each, while DPC++ used 16 blocks of size 1024. This analysis highlights the importance of work-group/item organisation, as well as the different choices the SYCL compilers make for this. 

\begin{figure}[t]
    \centering
    \includegraphics[width=\columnwidth]{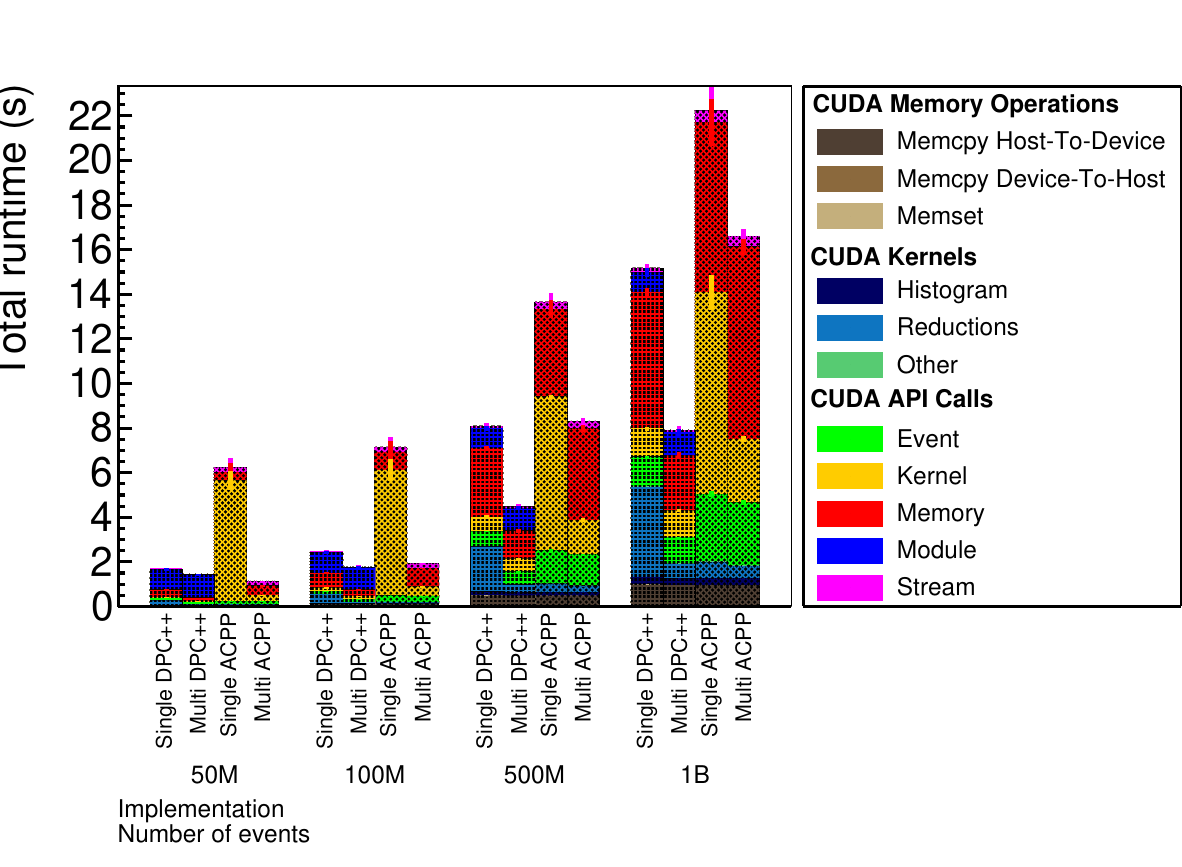}
    \caption{Total time spent on GPU activity in \texttt{Histo1D} with multiple reduction variables per SYCL kernel (multi) or a single reduction variable per kernel  (single). Two elements are reduced per work-item.}
    \label{fig:reduction_multi_vs_single_incl_api}
\end{figure}

To determine the performance benefit of combining our reductions into a single kernel, we benchmarked the difference in execution time with and without combining multiple reduction variables into a single SYCL kernel. \autoref{fig:reduction_multi_vs_single_incl_api} presents the result of this experiment, where the time spent on GPU activity is plotted against the number of events. The plot shows that we benefit from this optimisation in both SYCL implementations. We achieve a speedup of around 1.9x with DPC++ and 1.4x with AdaptiveCpp at one billion events. For DPC++, it is mostly the kernel runtime that improves, while for AdaptiveCpp, the kernel launch time improves. These results underline the benefits and importance of having easy-to-operate abstractions for common GPU programming patterns.

%%%%%%%%%%%%%%%%%%%%%%%%%%%%%%%%%%%%%%%%%%%%%%%%%%%%%%%%%%%%%%%%%%%%%%%%%%%%%%%%%%%%%%%%%%%%%%%%%%%%%%%%%%%%%%%%%
\subsection{Buffers vs. Device Pointers}\label{sec:buf_vs_ptr}
\noindent\fcolorbox{black}{ACMBlue!30}{
    \parbox{0.95\columnwidth}{%
\textbf{Insights gained:}
\begin{itemize}[leftmargin=*]
    \item Transferring a bulk of events using buffers performs similarly to using device pointers within the unified shared memory. 
    \item DPC++'s overall performance is closer to native CUDA than AdaptiveCpp, with both buffers and device pointers. 
\end{itemize}
}
}
\newline

As described in \autoref{sec:implementation}, we transfer bulks of event data from the host to the device in a loop. To achieve high performance, we want to minimise the memory transfer overhead as much as possible, using SYCL features for it. SYCL2020 provides several methods to manage allocations and data transfers. Most notable are the \textit{buffers/accessors} approach and \textit{Unified Shared Memory} (USM). There are three different USM allocations: for data exclusively located on the host (host pointers), exclusively on the device (device pointers), and for allocations with automatically managed transfers (shared pointers). 
To determine how effectively SYCL is optimizing data transfers, we select two strategies to compare: buffers with implicit transfers via accessors (BUF) and USM device pointers with explicit transfers (PTR). We expect the overhead of execution scheduling to be bigger with BUF, since the SYCL runtime needs to infer the necessary transfers, while for PTR they are already defined. 

\begin{figure}[!tbh]
    \centering
    \includegraphics[width=\columnwidth]{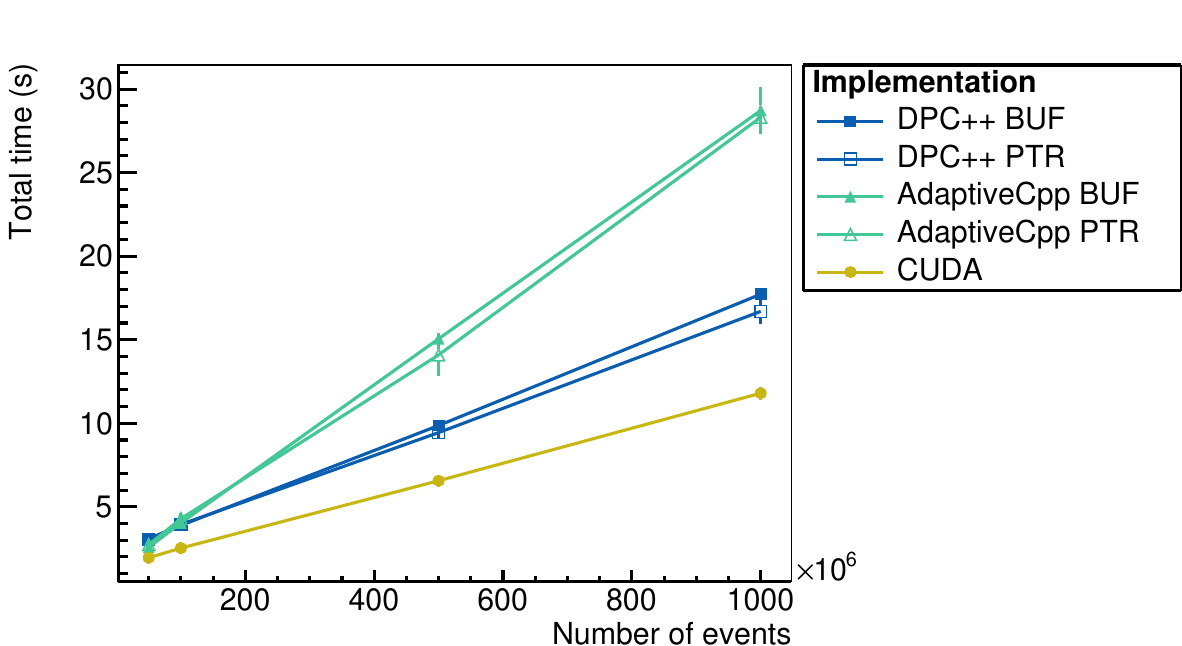}
    \caption{Average total runtime of \texttt{Histo1D} SYCL implementations with buffers and device pointers against CUDA.}
    \label{fig:buf_vs_usm_total}
\end{figure}

\autoref{fig:buf_vs_usm_total} compares the difference in total runtime measured with \texttt{steady\_clock} between DPC++, AdaptiveCpp, and native CUDA. The displayed runtimes include CPU time spent on reading in bulks, managing the RDataFrame computational loop, etc. 
With both AdaptiveCpp and DPC++, we do not observe a significant difference between the BUF and PTR versions. However, the DPC++ version performs much closer to the performance of native CUDA than the AdaptiveCpp version. With both compilers, the performance gap between the SYCL and CUDA implementations widens (that is, SYCL performance gets relatively worse) when the number of events increases. 

\begin{figure}[!tbh]
    \centering
    \includegraphics[width=\columnwidth]{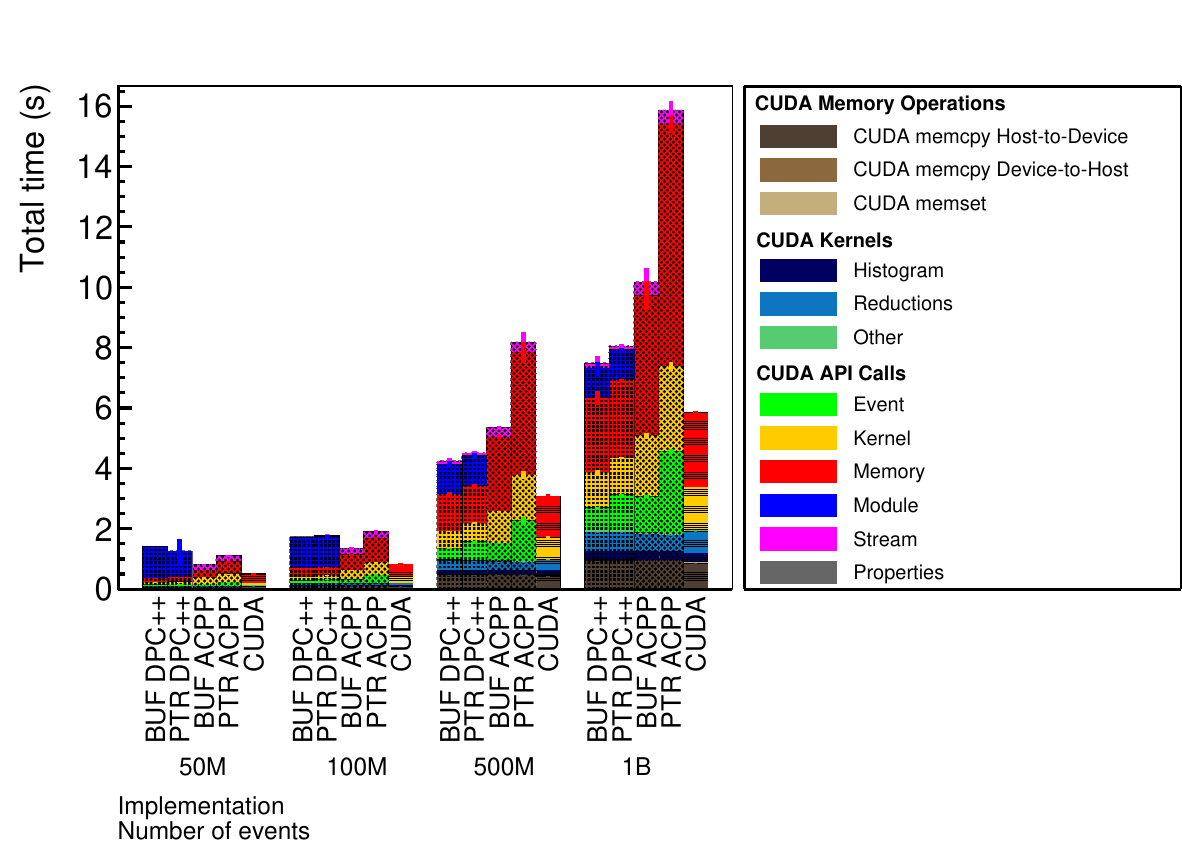}
    \caption{Total time spent on GPU activity in \texttt{Histo1D} with SYCL implementations, profiled using Nsight Systems.}
    \label{fig:buf_vs_usm_nsys}
\end{figure}

To explain these performance differences, we benchmark each version using NSight Systems. \autoref{fig:buf_vs_usm_nsys} shows the obtained results. Interestingly, more time is spent on GPU activity for PTR than for BUF, even though the total runtime was similar. This indicates that BUF incurs more overhead, which is not strictly related to GPU activity, on the CPU than PTR. The PTR versions spend approximately the same amount of time in kernels and memory operations as BUF. Since the number of computations and the amount of data transferred remains the same, when the number of events is identical, this is as expected.

When comparing the SYCL implementations with our native CUDA version, we see that the kernel runtimes are similar. However, the SYCL implementations have a larger overhead than native CUDA from API calls. An overhead visible with SYCL, but not with our CUDA version, is from streams and CUDA events. This is because the CUDA version used a single stream, while AdaptiveCpp created 4 streams for both PTR and BUF. The DPC++ implementation created significantly more: 67 streams for BUF and 192 for PTR, regardless of the number of events. We determined this by inspecting the number of calls to CUDA stream creation functions. Note that event API calls also include \texttt{cu(da)EventRecord} which is likely overhead from the profiler, to measure time intervals.

\begin{figure}[!thb]
    \centering
    \includegraphics[width=\columnwidth]{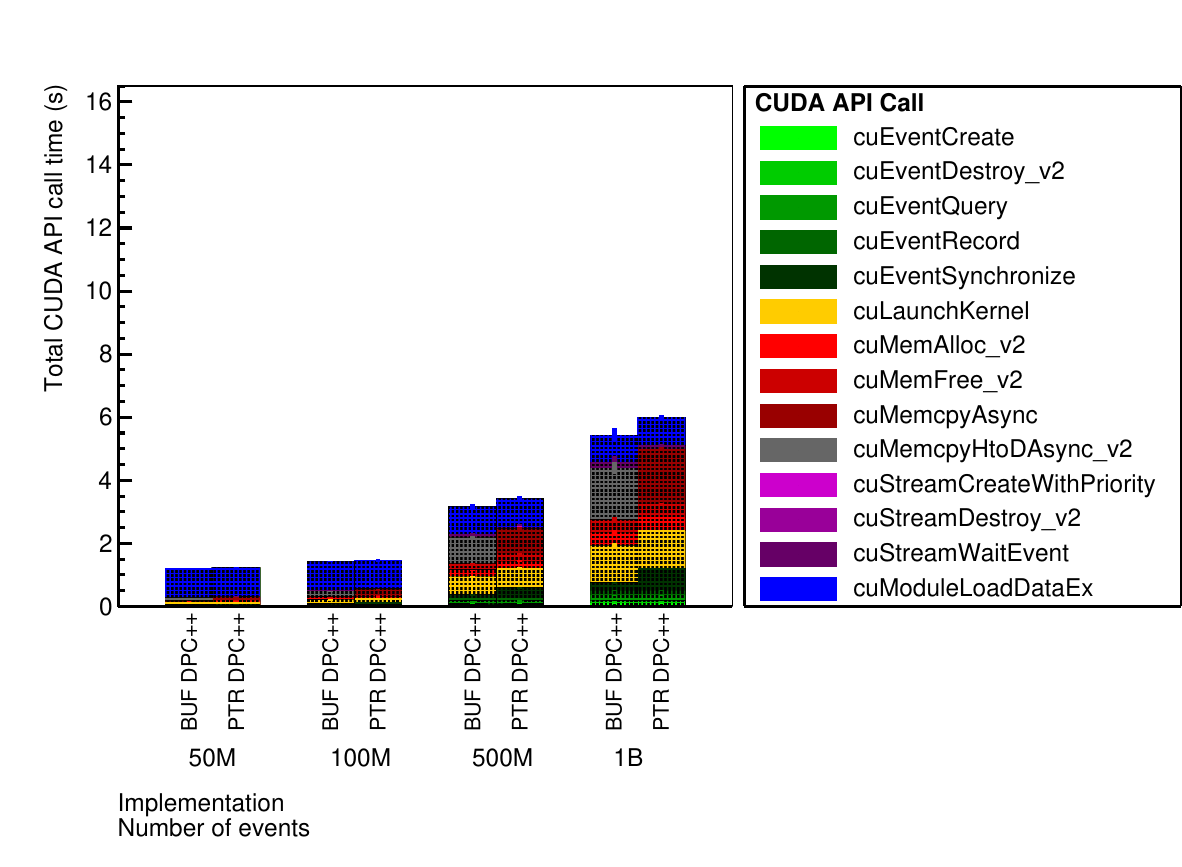}
    \caption{Total time spent on CUDA API calls in \texttt{Histo1D} with DPC++ featuring buffers (BUF) or device pointers (PTR), profiled using Nsight Systems.}
    \label{fig:dpcpp_buf_vs_ptr}
\end{figure}
\begin{figure}[!thb]
    \centering
    \includegraphics[width=\columnwidth]{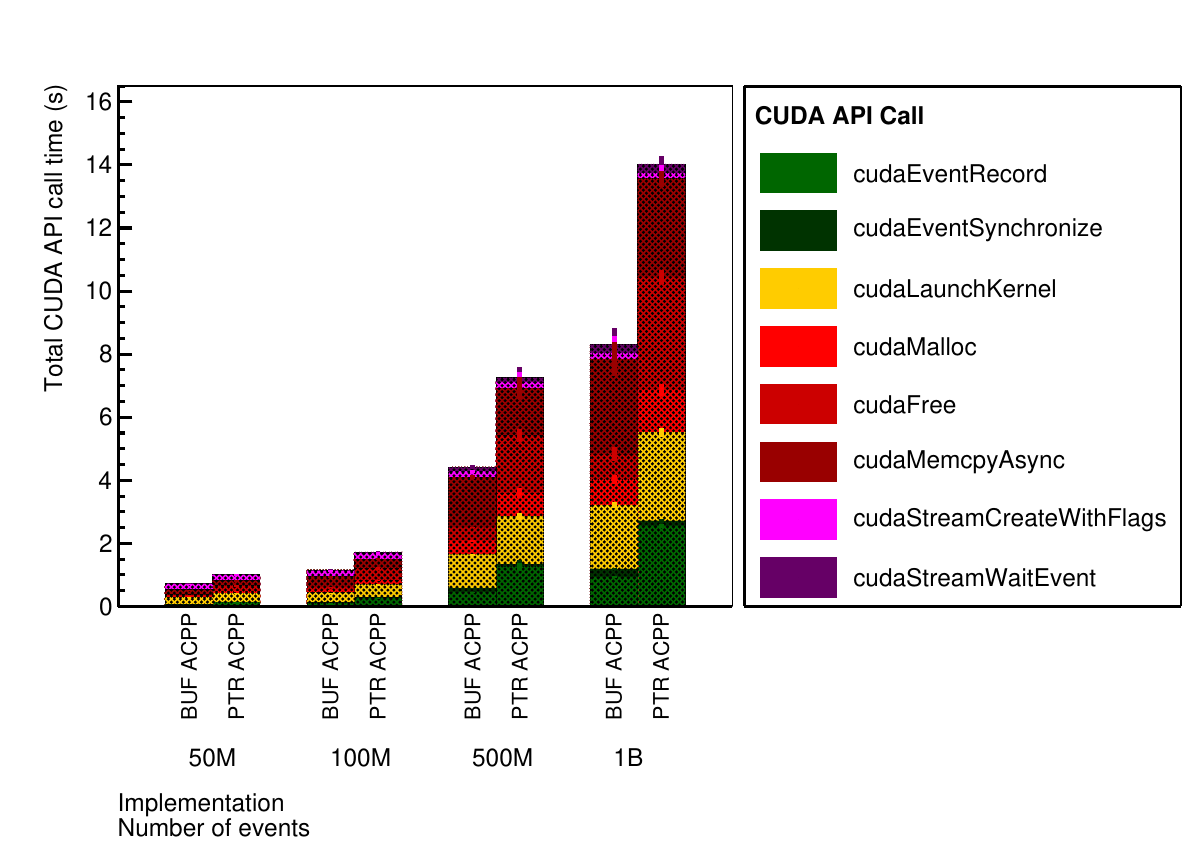}
    \caption{Total time spent on CUDA API calls in \texttt{Histo1D} with AdaptiveCpp featuring buffers (BUF) or device pointers (PTR), profiled using Nsight Systems.}
    \label{fig:acpp_buf_vs_ptr}
\end{figure}

In Figures \ref{fig:dpcpp_buf_vs_ptr} and \ref{fig:acpp_buf_vs_ptr}, we zoom into the specific CUDA API calls again. For DPC++, we see that version BUF has less overhead from event synchronisation than PTR, likely due to having fewer streams. This, combined with the observation that we have a minor difference in the total runtime, indicates that the scheduled execution was similar in both cases, but a non-negligible number of seemingly unnecessary streams were constructed.  
With AdaptiveCpp, there is a noticeable increase in runtime due to memory deallocations.
Moreover, AdaptiveCpp spent almost twice as much time on CUDA API calls than DPC++ for the same operations. 

Regarding usability, we had an easier time implementing dependencies using buffers+accessors than with device pointers. Using accessors, dependencies between command groups are implicitly defined by denoting the access mode (read, write, read+write). For device pointers, we need to define the dependencies manually. Debugging our dependency issues mainly involved a trial-and-error process of inserting additional synchronization points. With DPC++, the environment variable \texttt{SYCL\_PRINT\_ EXECUTION\_GRAPH} can be set to dump the execution graph to a DOT text file, which can be converted to a PNG image. However, documentation about the displayed elements is currently lacking. With AdaptiveCpp, the environment variable \texttt{ACPP\_DEBUG\_LEVEL=3} can be set to print dependency information along with other diagnostics, which can sometimes be hard to interpret due to the large amount of output. We believe that having easy-to-use dependency debugging tools would be extremely useful.

%%%%%%%%%%%%%%%%%%%%%%%%%%%%%%%%%%%%%%%%%%%%%%%%%%%%%%%%%%%%%%%%%%%%%%%%%%%%%%%%%%%%%%%%%%%%%%%%%%%%%%%%%%%%%%%%%
\subsection{Just-In-Time Compilation Overhead}\label{sec:aot_target}

% \begin{figure}[thb!]
\noindent\fcolorbox{black}{ACMBlue!30}{
    \parbox{0.95\columnwidth}{%
    \textbf{Insights gained:}
        \begin{itemize}[leftmargin=*]
            \item CUDA stores kernels in modules that are loaded on execution. If not all kernels in a CUDA module are used, some kernels will be loaded unnecessarily, causing extra overhead.
            \item By default, AdaptiveCpp creates a CUDA module for each kernel, while DPC++ creates a single module for all kernels.
            % \item Enabling ahead-of-time compilation of SYCL code can significantly improve the runtime performance.
        \end{itemize}
    }%
}
\newline

\noindent\fcolorbox{black}{ACMBlue!30}{
    \parbox{0.95\columnwidth}{%
        \begin{itemize}[leftmargin=*]
            \item Enabling ahead-of-time compilation of SYCL code can significantly improve the runtime performance.
        \end{itemize}
    }%
}
\newline

Both AdaptiveCpp and DPC++ expect a target specification when invoking their compiler for ahead-of-time compilation. When targeting CUDA, the architecture/compute capability also needs to be specified. While developing our SYCL versions, we noticed a large performance gap between execution on different NVIDIA GPUs, which we did not observe with our native CUDA version. Upon further investigation using NSight Systems, we discovered that this was due to an incorrect target compute capability setting, resulting in a higher just-in-time compilation overhead due to incompatible code. Despite the incorrect option, the program still executed correctly. 

To illustrate the performance benefit of ahead-of-time compilation, we benchmarked our application with the following settings: targeting a lower compute capability (\texttt{cuda:sm\_75}) and targeting the actual architecture of our NVIDIA RTX A4000 GPU (\texttt{cuda:sm\_86}).
Since there is no difference in the execution time of the kernels, we only show the results of the time spent in CUDA API calls. To ensure the readability of the legends, we omit the API calls for which the percentage of the time, relative to the total time spent on API calls, is negligible ($<1.0\%$).

\begin{figure}[!tbh]
    \centering
    \includegraphics[width=\columnwidth]{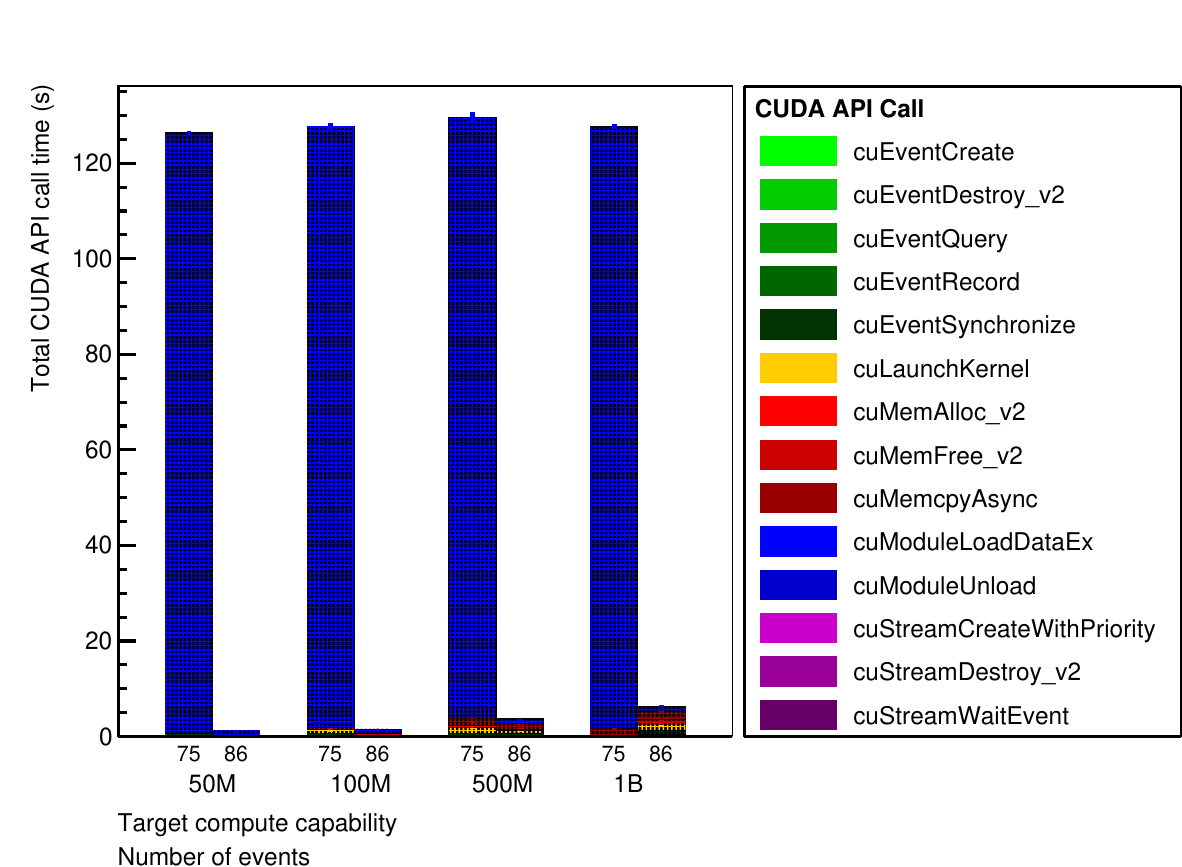}
    \caption{Total time spent on CUDA API calls in \texttt{Histo1D} with DPC++ compilation targeting either cuda:sm\_75 (incompatible) or cuda:sm\_86 (compatible).}
    \label{fig:dpcpp_aot_vs_jit}
\end{figure}

The results for DPC++ are shown in 
\autoref{fig:dpcpp_aot_vs_jit}. We immediately notice that a significant amount of time is spent on the CUDA API call \texttt{cuModuleLoadDataEx} for the version targeting a lower compute capability. This API call is related to the compilation of device code. The CUDA driver recognises kernels written either in assembly form (\textit{PTX}) or in cuda binary form (\textit{cubin}) \cite{cuda_programming_guide}. If the code is compiled into PTX, the CUDA driver needs to compile it \textit{just-in-time} (JIT) into cubin at runtime before it can be executed. If the code was already compiled into cubin \textit{ahead-of-time} (AOT), this step is not necessary. A CUDA module refers to dynamically loadable packages of device code and data, in PTX or cubin form. 

\begin{figure}[!tbh]
    \centering
    \includegraphics[width=\columnwidth]{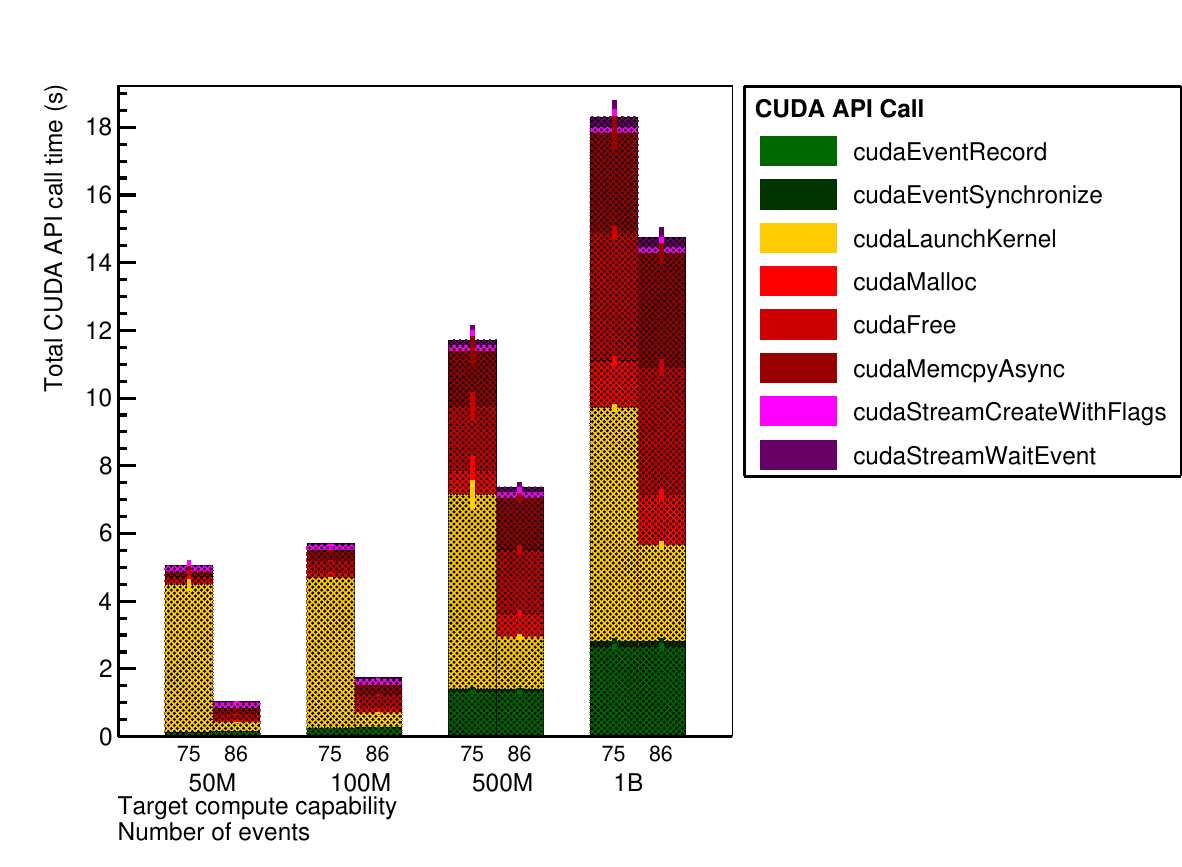}
    \caption{Total time spent on CUDA API calls in \texttt{Histo1D} with AdaptiveCpp compilation targeting cuda:sm\_75 (incompatible) or cuda:sm\_86 (compatible).}
    \label{fig:acpp_aot_vs_jit}
\end{figure}

The large \texttt{cuModuleLoadDataEx} overhead, therefore, suggests that the kernel code with an incompatible compute capability was JIT compiled. The overhead incurred by this method causes a downgrade in performance up to 98x in the case of 500 million events. The DPC++ clang compiler compiles the SYCL CUDA code into PTX and then assembles the resulting file into cubin at link time \cite{dpcpp_compilation}. Both the PTX file and cubin code are assembled into a CUDA \textit{fat binary} (fatbin). Thus, when a kernel is launched, the CUDA runtime first looks for a compatible cubin, otherwise it JITs the PTX file. While PTX leads to longer (initial) load times, it is forward-compatible with major and minor revisions of the compute capability, while cubin code only remains executable with minor revisions \cite{cuda_programming_guide}. This explains why our code was still functioning correctly despite targeting an older architecture. 

To decrease the overhead of JITting, the CUDA driver is capable of caching JITted cubin code to avoid recompilation on subsequent runs of the application \cite{cuda_jit_fatbin}. For this reason, we did not observe the same overhead with our native CUDA version when switching compute capabilities. In the case of DPC++, it seems that the cubin code could not be (fully) cached. Even when targeting the correct compute capability, an overhead of roughly 1 second from \texttt{cuModuleLoadDataEx} remains. In \autoref{fig:dpcpp_buf_vs_ptr}, we see that the performance of DPC++ matches the performance of our native CUDA version without this overhead. Possible causes could be that the cache was full or that (non-existent) modifications were detected, which results in re-JITting every run. Further investigation is needed to determine the exact cause. 

The results with AdaptiveCpp are shown in \autoref{fig:acpp_aot_vs_jit}, where the colours for each API call match the ones with similar functionality in \autoref{fig:dpcpp_aot_vs_jit}. Compared to DPC++, the impact of targeting the incorrect compute capability is relatively low, with a difference of only up to 4.9x (for 50M events). Similarly to DPC++, AdaptiveCpp generates PTX code using the clang CUDA toolchain. Here, we do not observe a large overhead caused by \texttt{cuModuleLoadDataEx}. After an investigation by the Intel DPC++ developers, we discovered that this is due to a different kernel-module splitting granularity. AdaptiveCpp splits each kernel into a separate CUDA module by default, while DPC++ prefers to assemble all kernels into a single module. In our case-study, we only execute one of the 60 instantiations of our templated kernel, so many kernels are loaded unnecessarily in the case of DPC++. When modifying the splitting behavior to match using \texttt{-fsycl-\allowbreak{}device-\allowbreak{}code-split=per\_kernel}, we get similar results for DPC++ as with AdaptiveCpp. 

% Instead, only the \texttt{cudaLaunchKernel} API call is affected. The increased kernel launch duration with compute capability 75 also appeared in the DPC++ execution. 

% An interesting difference we noticed in our NSight Systems profiling results is that AdaptiveCpp uses CUDA's Runtime API, while DPC++ uses the Driver API. This can be determined by the API call names, which start with either \texttt{cu} or \texttt{cuda} for the Runtime API and the Driver API, respectively. The driver API is lower level and allows for fine-grained control over context management and module management \cite{cuda_runtime_driver_api}. With the runtime API, this is done implicitly. Thus, the difference in caching between AdaptiveCpp and DPC++ could be due to a difference in context and module management. 

\subsection{Implementation Advice}
Based on our analysis and findings, we provide several guidelines for developers aiming to implement their own SYCL code, either from scratch or as a migration from native CUDA code. These suggestions stem from the challenges we faced during implementation and the insights highlighted in the previous sections. 
\begin{itemize}[leftmargin=*]
    \item Try out multiple SYCL compiler implementations. This helps with discovering bugs caused by assumptions about undefined behaviour in the specification. 
    \item If a program has unexpected results, try to insert additional synchronisation points, to check that the enqueued work is executed in the expected order. Alternatively, set the environment variable \texttt{SYCL\_PRINT\_EXECUTION\_GRAPH} when running the application with DPC++, to output the dependency graph. For AdaptiveCpp, \texttt{ACPP\_DEBUG\_LEVEL=3} can be set to print dependency debugging information. 
    \item When using SYCL2020 reductions, experiment with increasing the amount of work assigned to each work-item and the number of reduction variables per kernel.
    \item Experiment with different granularities for splitting kernels into modules to minimize the overhead of loading kernel modules. For example, with DPC++, this involves modifying the \texttt{-fsycl-\allowbreak{}device-\allowbreak{}code-split.}
\end{itemize}
\section{Related Work}\label{sec:related}
To the best of our knowledge, ours is the first SYCL vs. CUDA study in the context of ROOT and its histrogram operation. However, other SYCL comparative performance studies do exist. 

For example, also in high-energy physics, \citeauthor{Joube_2023} compare the performance and ease-of-programming of SYCL2020 data transfers using buffers and USM (host, device, and shared pointers) \cite{Joube_2023}. They applied both approaches to the particle tracking library, Traccc, and document the challenges and differences in the context of this application. 

\citeauthor{baratta2022} analyse the performance of SYCL2020 reductions with similar optimisations in the context of a linear system solver in both AdaptiveCpp and DPC++, compared to CUDA BLAS \cite{baratta2022}. They also found that increasing the workload per work-item and fusing multiple reductions into a single kernel greatly increases the performance of the built-in reduction for AdaptiveCpp. Overall, their results show that AdaptiveCpp's reduction performs reasonably, while DPC++ performs poorly (at small problem sizes). Since SYCL2020 reduction support is quite recent, most works investigating reduction semantics in SYCL implemented their own kernel \cite{jin2021, jin2022, deakin2020}.

Overall, we believe our performance analysis study is similar in spirit with such studies, but the application, the analysis methodology, and the findings differ given our choice of compilers, the hardware platform, and our specific approach.

\section{Conclusion}\label{sec:conclusion}
In this work, we described our experiences with porting a core high-energy physics data-analysis operation from the ROOT framework, \textit{histogramming}, from native CUDA to SYCL. Our implementations successfully target both AdaptiveCpp and DPC++. In a comparative analysis, we investigated three SYCL specific features: (1) the usage of SYCL2020 reductions, (2) the choice for data transfers using buffers versus device USM pointers, and (3) the impact of just-in-time PTX compilation caching. We discovered that the performance of the built-in reduction kernels can be improved significantly by increasing the workload assigned to each work-item and combining separate reductions into a single kernel. While we did not observe a significant difference in the total runtime between SYCL buffers+accessors compared to USM device pointers, we noticed a trade-off between the CPU and GPU overheads. Furthermore, our results show that DPC++ has a higher overhead from just-in-time compilation of generated PTX code than AdaptiveCpp, due to less efficient caching. Overall, our results indicate that DPC++ performs considerably better than AdaptiveCpp, and it is closer to (albeit still lower than)  the performance of our native CUDA implementation. 

On the methodological side, we recommend combining multiple diagnostic tools. We found that benchmarking a SYCL CUDA application with NSight Systems can give a good insight into performance bottlenecks: it is not sufficient to analyse performance solely based on kernel runtimes; instead, the statistics from CUDA API calls should be taken into account as well to fully understand performance differences. 

Although our analysis indicates several “handle-with-care" parts of SYCL programming and some performance pitfalls, we note that both SYCL implementations, AdaptiveCpp and Intel DPC++, are still under active development. Therefore, it is necessary for users to regularly re-investigate the performance of these compilers, as their behaviour might further improve in the future. 

Currently, for our ROOT use case, SYCL does not outperform CUDA. The application has small computational kernels and large kernel launch and memory transfer overheads. Additionally, the kernels contain a non-negligible amount of branching to handle different histogram parameters - e.g., different dimensions or input data types. This combination of factors leads to suboptimal performance of the application in CUDA, which is further exacerbated when using SYCL due to the additional overhead of managing the runtime. However, we still gain several benefits from using SYCL. Firstly, we acquired support for accelerators beyond NVIDIA GPUs and parallelism within bulks on the CPU - before SYCL, RDataFrame only supported multithreaded execution across bulks. Moreover, thanks to abstractions such as the SYCL2020 reduction interface, our code has become more concise and descriptive compared to our CUDA implementation.

In future work, we want to experiment with different accelerator architectures using our SYCL implementations. We want to test the compatibility and performance on accelerators of vendors other than NVIDIA, such as Intel and AMD. Additionally, we want to benchmark more complex ROOT use cases. Instead of having a single histogram action, we could compute multiple histograms using data in different (parts of the) columns, which results in more complicated data dependencies. Moreover, it is worth investigating the migration of more RDataFrame actions to the GPU and pipelining the processing of bulks to increase the overlap of communication with computation.

% \begin{description}
% \item[\texttt{sidebar}:]  Place formatted text in the margin.
% \item[\texttt{marginfigure}:] Place a figure in the margin.
% \item[\texttt{margintable}:] Place a table in the margin.
% \end{description}

%%
%% The acknowledgments section is defined using the "acks" environment
%% (and NOT an unnumbered section). This ensures the proper
%% identification of the section in the article metadata, and the
%% consistent spelling of the heading.
\begin{acks}
Partial support for this work was provided by the Horizon Europe Project ``SYCLOPS", funded from the European Union HE Research and Innovation programme under grant agreement No 10109287. 
\end{acks}
% 

%%
%% Print the bibliography
%%
\printbibliography
\vfill\null
%\newcolumn
%%
%% If your work has an appendix, this is the place to put it.
\appendix\label{sec:appendix}
\section{Reproducibility}
In this section, we describe the steps that are needed to reproduce the experiments performed in this paper in more detail. The ROOT code is available on GitHub \cite{gpu_histogram_bulk}. To build this development branch, the following CMake command can be executed: 

\begin{verbatim}
    cmake  /path/to/root-source 
    -DCMAKE_BUILD_TYPE=Release 
    -DCMAKE_CXX_STANDARD=17 -Dcuda=ON 
    -DCMAKE_CUDA_ARCHITECTURES=<cuda_arch> 
    -DCUDA_TOOLKIT_DIR=</path/to/cuda-toolkit> 
    -DCMAKE_CUDA_COMPILER=<path/to/nvcc> 
    -Doneapi=OFF -Dadaptivecpp=ON 
    -DAdaptiveCpp_DIR=/path/to/AdaptiveCpp/lib/cmake
    -DACPP_TARGETS="cuda:<arch>"
\end{verbatim}

To build with DPC++ enabled, the CMake command below is required. Note that AdaptiveCpp needs to be disabled to enable DPC++ and vice versa. 
\begin{verbatim}
    cmake  /path/to/root-source 
    -DCMAKE_BUILD_TYPE=Release 
    -DCMAKE_CXX_STANDARD=17 -Dcuda=ON 
    -DCMAKE_CUDA_ARCHITECTURES=<cuda_arch> 
    -DCUDA_TOOLKIT_DIR=</path/to/cuda-toolkit> 
    -DCMAKE_CUDA_COMPILER=<path/to/nvcc> 
    -Doneapi=ON -Dadaptivecpp=OFF 
    -DSYCL_INCLUDE_DIR=/path/to/DPC++/include 
    -DSYCL_LIB_DIR=/path/to/DPC++/lib
    -DSYCL_COMPILER=/path/to/DPC++/bin/clang++
\end{verbatim}
When using a DPC++ version that is built from source, it is important that an installation of OpenCL is available on the system to target the CPU. The environment variable \cs{OCL\_ICD\_FILENAMES} needs to be set to the path where \cs{libintelocl.so} can be found.

\end{document}